\documentclass[conference]{IEEEtran}
\IEEEoverridecommandlockouts

\usepackage[hidelinks]{hyperref}

\usepackage{tikz}
\usetikzlibrary{shapes.geometric, arrows.meta, positioning, fit, backgrounds}
\usepackage{booktabs}
\usetikzlibrary{calc} 

\usepackage{cite}
\usepackage{amsmath,amssymb,amsfonts}
\usepackage{algorithmic}
\usepackage{graphicx}
\usepackage{textcomp}
\usepackage{xcolor}
\usepackage{randtext} 
\usepackage{verbatim}

\def\BibTeX{{\rm B\kern-.05em{\sc i\kern-.025em b}\kern-.08em
    T\kern-.1667em\lower.7ex\hbox{E}\kern-.125emX}}
\begin{document}

\title{Vendor-Aware Industrial Agents: RAG-Enhanced LLMs for Secure On-Premise PLC Code Generation}

\author{\IEEEauthorblockN{Joschka Kersting}
\IEEEauthorblockA{\textit{Centre for Machine Learning}\\
\textit{Fraunhofer IOSB-INA} \\
Lemgo, Germany \\
\randomize{joschka.kersting@iosb-ina.fraunhofer.de}}
\and
\IEEEauthorblockN{Michael Rummel}
\IEEEauthorblockA{\textit{FA-EDC} \\
\textit{Mitsubishi Electric Europe }\\Ratingen, Germany \\
\randomize{michael.rummel@meg.mee.com}
\and
\IEEEauthorblockN{Gesa Benndorf}
\IEEEauthorblockA{\textit{Centre for Machine Learning}}
\textit{Fraunhofer IOSB-INA} \\
Lemgo, Germany \\
\randomize{gesa.benndorf@iosb-ina.fraunhofer.de}}
}

\maketitle

\begin{abstract}
Programmable Logic Controllers are operated by proprietary code dialects; this makes it challenging to train coding assistants. Current LLMs are trained on large code datasets and are capable of writing IEC 61131-3 compatible code out of the box, but they neither know specific function blocks, nor related project code. Moreover, companies like Mitsubishi Electric and their customers do not trust cloud providers. Hence, an own coding agent is the desired solution to cope with this. In this study, we present our work on a low-data domain coding assistant solution for industrial use. We show how we achieved high quality code generation without fine-tuning large models and by fine-tuning small local models for edge device usage. Our tool lets several AI models compete with each other, uses reasoning, corrects bugs automatically and checks code validity by compiling it directly in the chat interface. We support our approach with an extensive evaluation that comes with code compilation statistics and user ratings. We found that a Retrieval-Augmented Generation (RAG) supported coding assistant can work in low-data domains by using extensive prompt engineering and directed retrieval. 
\end{abstract}


\section{Introduction}

Every day, millions of Programmable Logic Controllers (PLCs) orchestrate production lines, manage power grids, and safeguard critical infrastructure—from automotive assembly plants to water treatment facilities \cite{John2010}. Behind each automated process lies control logic programmed in specialized languages, most prominently Structured Text (ST), a high-level language standardized in IEC 61131-3 that combines the expressiveness of modern programming with the real-time determinism industrial systems demand \cite{John2010}.

Despite ST's syntactic similarity to languages like Python or Pascal, developing PLC control logic remains fundamentally different. Where Python developers access thousands of open-source libraries and Stack Overflow answers, ST engineers face sparse code repositories and proprietary vendor documentation. Moreover, each manufacturer -- Siemens, CODESYS, Mitsubishi Electric, Rockwell Automation -- implements distinct dialects with incompatible function blocks, naming conventions, and compiler behaviors \cite{Yang2024,Haag2024}. A timer implementation for Siemens TIA Portal will not compile on a Mitsubishi Electric iQ-R controller; even seemingly portable constructs like counter instructions exhibit vendor-specific reset semantics and enable/disable conventions.

This fragmentation collides with industrial realities: manufacturing firms guard control logic as trade secrets, regulatory frameworks restrict proprietary code sharing, and security-critical infrastructure operators prohibit cloud-dependent AI tools that could expose operational intelligence \cite{Brehme2025}. The result is a low-data domain where traditional large language model training approaches—reliant on vast public code corpora—struggle to generate vendor-compliant, compilable control logic \cite{Haag2024}. Existing coding assistants trained on GitHub's predominantly web and enterprise software fail to recognize that \texttt{RTRIG\_P} triggers on rising edges in Mitsubishi Electric ST, that variable declarations must occur in external label editors rather than inline \texttt{VAR} blocks, or that certain ladder logic instructions are forbidden in ST contexts.

Nonetheless, recent research has demonstrated significant advances in automated ST code generation using Large Language Models (LLMs) \cite{Xia2025}. Haag et al. \cite{Haag2024} pioneered the integration of compiler feedback with LLM training, achieving 70\% compilation success rates through iterative Direct Preference Optimization (DPO). Liu et al. \cite{Liu2024} developed Agents4PLC, achieving up to 68.8\% formal verification success across 23 programming tasks. Yang et al. \cite{Yang2024} advanced the field with AutoPLC, a framework for vendor-aware ST code generation achieving 60\% compilation success across Siemens TIA Portal and CODESYS platforms. Meanwhile, Ren et al. \cite{Ren2025} introduced MetaIndux-PLC, employing control logic-guided iterative fine-tuning.

However, despite these promising academic results, significant gaps remain between research achievements and industrial deployment requirements \cite{Brehme2025}. Brehme et al.’s \cite{Brehme2025} comprehensive industry study revealed that most RAG applications in industrial settings remain in prototype stages, with data protection, security, and quality identified as paramount concerns (8.9, 8.5, and 8.7 of 10 points).

Existing research has primarily focused on either general IEC 61131-3 compatibility or multiple vendor platforms simultaneously, but has not adequately addressed the challenge of creating practical coding assistants for specific proprietary implementations \cite{Brehme2025}. While Haag et al. \cite{Haag2024} demonstrated compiler feedback integration, their approach requires extensive fine-tuning and relies on cloud-based services. Yang et al. \cite{Yang2024} addressed multi-vendor variations but depends on cloud-based processing incompatible with industrial security requirements \cite{Brehme2025}. Liu et al. \cite{Liu2024} operates as a single-model solution lacking comprehensive retrieval mechanisms for proprietary function blocks. Ren et al. \cite{Ren2025} requires substantial fine-tuning datasets typically unavailable in proprietary environments.

Recent work has extended LLM-based control logic generation beyond 
textual notations. Koziolek et al. \cite{Koziolek2025} 
demonstrated Spec2Control, a workflow generating graphical DCS control logic for ABB systems with a high connection accuracy across 65 test cases, achieving 94-96\% labor savings. While Spec2Control targets multi-plant DCS graphical programming, our work focuses on vendor-specific textual ST generation.

The specialized nature of industrial automation creates a low-data domain scenario where comprehensive training datasets for specific vendor implementations are scarce or confidential \cite{Haag2024}, making traditional fine-tuning approaches impractical \cite{Brehme2025}. Notably, no existing work has specifically addressed Mitsubishi Electric's ST implementation and MELSOFT GX\,Works3, the engineering environment for Mitsubishi Electric's PLC series,  which presents challenges due to its proprietary function blocks, specific syntax variations, and integration requirements. These are, however, compliant with the IEC 61131-3 standard, since it leaves freedoms to vendors. These freedoms and said function blocks, are not known to the wider public and hence not known to LLMs.

This work addresses this critical gap through a novel RAG-enhanced coding assistant specifically designed for Mitsubishi Electric’s ST implementation with the development environment and compiler MELSOFT GX\,Works3. Unlike previous approaches targeting multiple vendors \cite{Yang2024} or require extensive fine-tuning \cite{Haag2024,Ren2025}, our solution focuses on one ecosystem.

\begin{figure}[h!t]
    \centering
    \includegraphics[width=1\linewidth]{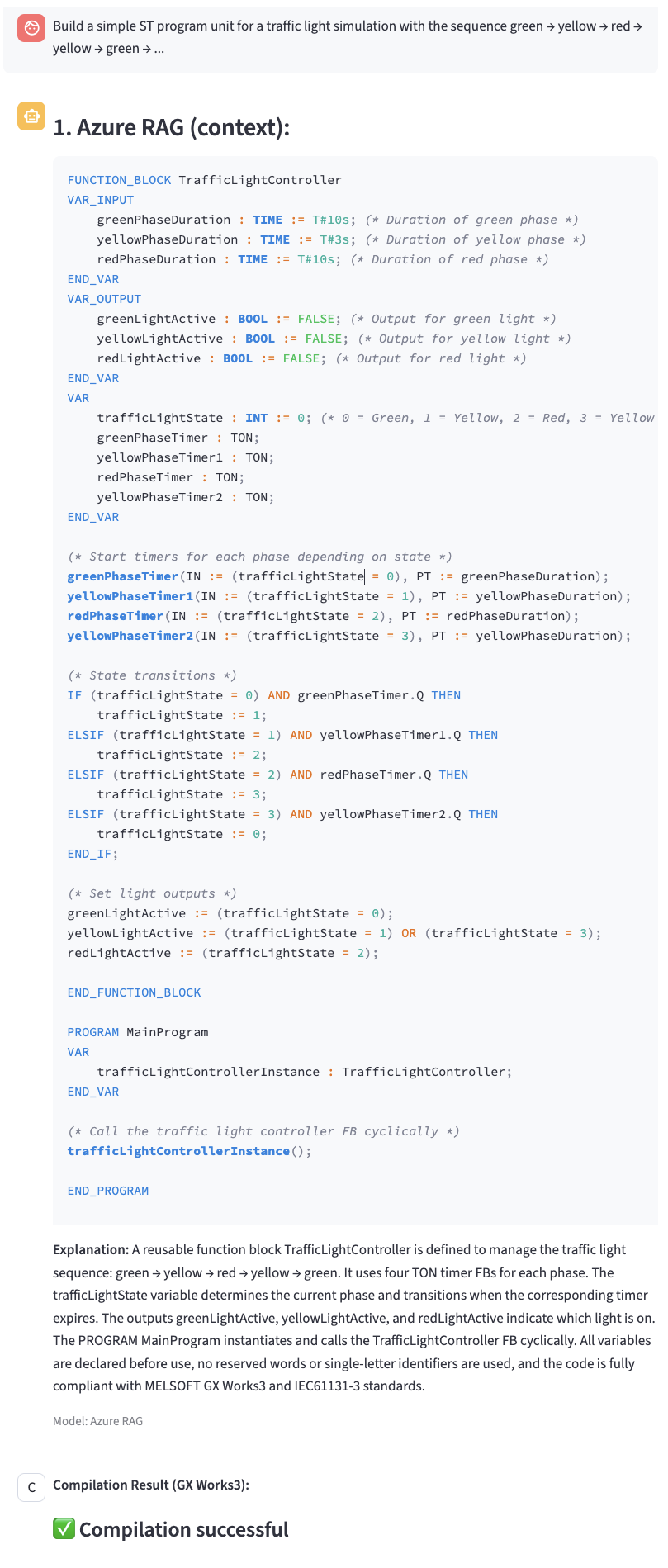}
    \caption{Example of generated ST code using our GeneriST web app (built with Streamlit \cite{streamlit2025}).}
    \label{fig:generation_example}
    \vspace{1cm}
\end{figure}

The key contributions include: (1) the first RAG-based coding assistant specifically designed for Mitsubishi Electric’s ST implementation, addressing a significant gap in vendor-specific industrial AI tools; (2) novel competitive multi-model approach with MELSOFT GX\,Works3 integration enabling real-time compilation feedback and error correction; (3) comprehensive evaluation using Mitsubishi Electric’s development environment with both compilation statistics and expert assessments from engineers familiar with the platform; and (4) practical demonstration that RAG-supported coding assistants can achieve high-quality code generation for specific proprietary platforms through targeted knowledge base construction and vendor-specific prompt engineering. We also enhance earlier work \cite{Reinhardt2025} that used translation transformers due to privacy constraints and plan to further support local Small and Medium Sized companies (SMEs) in the future with our results. Fig.~\ref{fig:generation_example} illustrates an ST program generated by our approach.

\section{Industrial Project Background and~System~Architecture}


GX\,Works3 emphasizes a label\text- and device\text-based model in which variables (labels) are defined and managed in the engineering tool and then referenced by ST programs; the operating manual details dedicated editors and the standard workflow for creating ST programs, running syntax checks, and converting projects \cite{MEC2024}.

The ST programming guide presents a program creation procedure that explicitly ties label definition, ST authoring, and conversion (compilation) into an executable sequence program, including correction steps when compile errors occur \cite{MEC2022}. Mitsubishi Electric instructions are exposed as functions and function blocks callable from ST \cite{MEC2023,MEC2024b}.

Accordingly, the project scope is a specific copilot that integrates with GX\,Works3 workflows: it uses directed retrieval over official documentation and internal code patterns to emit code that compiles in GX\,Works3. Moreover, the copilot validates outputs via the platform’s compilation and syntax-check mechanisms \cite{MEC2022,MEC2024}.

Finally, under low-data constraints, synthetic and customer-derived task variants are iteratively refined using compiler feedback, aligning with evidence that compilation-aware feedback loops yield measurable improvements in ST code generation quality \cite{Haag2024}.


\begin{figure}[t]
\centering
\begin{tikzpicture}[scale=0.77, transform shape,
    node distance=0.8cm and 1.2cm,
    block/.style={rectangle, draw, fill=blue!10, text width=3cm, text centered, rounded corners, minimum height=1cm, font=\small},
    datablock/.style={rectangle, draw, fill=green!10, text width=2.5cm, text centered, rounded corners, minimum height=0.8cm, font=\footnotesize},
    modelblock/.style={rectangle, draw, fill=orange!10, text width=2.8cm, text centered, rounded corners, minimum height=0.8cm, font=\footnotesize},
    compilerblock/.style={rectangle, draw, fill=red!10, text width=2.5cm, text centered, rounded corners, minimum height=0.8cm, font=\footnotesize},
    arrow/.style={-Stealth, thick},
    dashedarrow/.style={-Stealth, thick, dashed}
]

\node[datablock] (fb) {Function Blocks\\(iQ-R CPU)};
\node[datablock, right=of fb] (specs) {Specifications\\(MEC2023/24)};
\node[datablock, right=of specs] (misc) {Auxiliary Context\\(Uploads, Chats)};

\node[block, below=1.5cm of specs] (retrieval) {Knowledge Retrieval\\3 Specialized Retrievers\\(Libraries, Specs, General)};

\node[block, left=1cm of retrieval] (user) {User Query};
\node[block, below=0.5cm of user] (expansion) {Optional Query\\Expansion\\(LLM Reasoning)};

\node[block, below=1.5cm of retrieval] (prompt) {Prompt Construction\\Hard Constraints + Retrieved Context\\+ POU Templates + Canonical Example};

\node[modelblock, below left=1.5cm and 0.5cm of prompt] (azure_rag) {Azure RAG\\GPT-4.1\\+ Retrieval};
\node[modelblock, below=1.5cm of prompt] (azure_std) {Azure Standard\\GPT-4.1\\No Retrieval};
\node[modelblock, below right=1.5cm and 0.5cm of prompt] (local_rag) {Local RAG\\Fine-tuned GGUF\\+ Retrieval};

\node[block, below=2cm of azure_std] (codegen) {Generated ST Code\\(JSON or Markdown)};

\node[compilerblock, below=1.2cm of codegen] (compiler) {GX\,Works3\\Compilation Server\\(MCP)};

\node[block, right=0.8cm of compiler] (repair) {Diagnostic-Guided\\Repair\\(Max 3 Iterations)};

\node[block, below=1.2cm of compiler] (output) {Validated ST Code\\+ Compilation Report};

\node[block, left=0.7cm of output] (ui) {User Interface\\Model Selection, Upload,\\Chat Persistence};

\draw[arrow] (fb) -- (retrieval);
\draw[arrow] (specs) -- (retrieval);
\draw[arrow] (misc) -- (retrieval);
\draw[arrow] (user) -- (expansion);
\draw[arrow] (expansion) -- (retrieval);
\draw[arrow] (retrieval) -- (prompt);
\draw[arrow] (prompt) -- (azure_rag);
\draw[arrow] (prompt) -- (azure_std);
\draw[arrow] (prompt) -- (local_rag);
\draw[arrow] (azure_rag) -- (codegen);
\draw[arrow] (azure_std) -- (codegen);
\draw[arrow] (local_rag) -- (codegen);
\draw[arrow] (codegen) -- (compiler);
\draw[arrow] (compiler) -- (output);
\draw[arrow] (ui) -- (output);

\draw[dashedarrow] (compiler) -- (repair);

\begin{scope}[on background layer]
\draw[dashedarrow] (repair) |- node[near end, above, font=\small] {Failure} (prompt);
\end{scope}

\draw[dashedarrow] (compiler) -- node[near end, right, font=\small] {Success} (output);

\begin{scope}[on background layer]
\node[draw, dashed, fit=(fb)(specs)(misc), fill=green!5, inner sep=0.3cm, label=above:{\textbf{Knowledge Base}}] {};
\node[draw, dashed, fit=(azure_rag)(azure_std)(local_rag), fill=orange!5, inner sep=0.3cm, ] (cm) {};
\node[draw, dashed, fit=(compiler)(repair), fill=red!5, inner sep=0.25cm, label=above:{\textbf{Compiler-in-the-Loop}}] {};
\end{scope}

\node[fill=white, inner sep=2pt, rounded corners=0.25pt] at ($(cm.north)+(0, 6.33pt)$) {\textbf{Competitive\,Model\,Layer}};

\end{tikzpicture}
\caption{System architecture for Mitsubishi Electric ST coding assistant with RAG, multi-model orchestration, and compiler-driven iterative repair.}
\label{fig:architecture}
\end{figure}
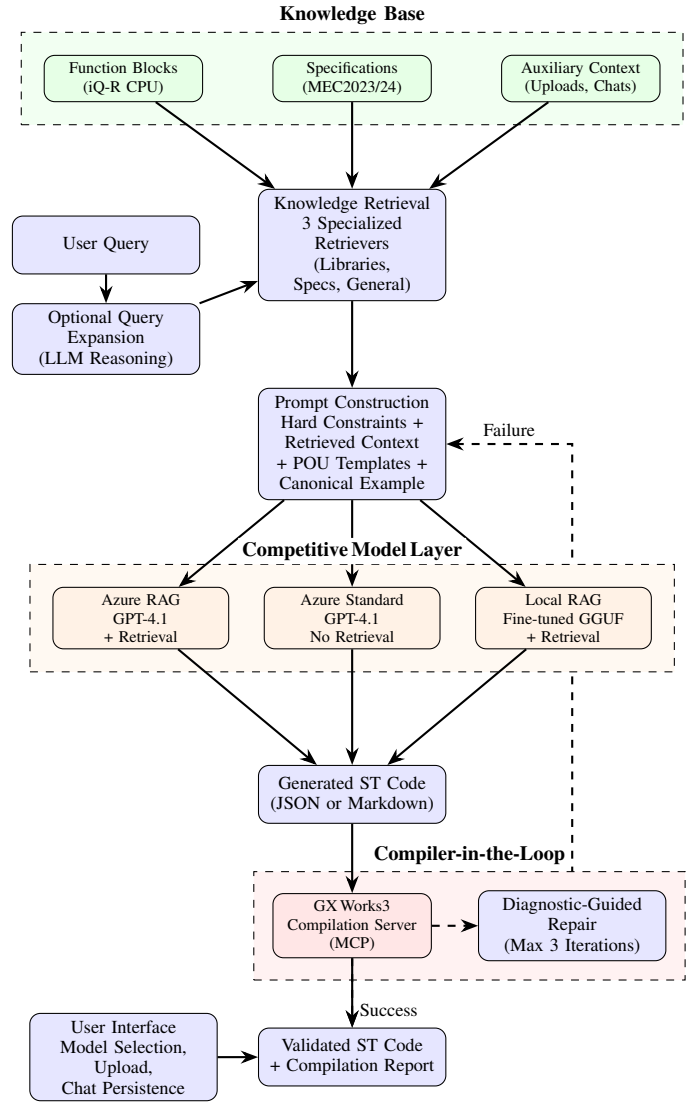


The system architecture is determined by a RAG pipeline that combines directed knowledge retrieval, competitive multi\text-model orchestration, and compiler\text-in\text-the\text-loop validation specifically for Mitsubishi Electric's iQ\text-R ST dialect (Fig.~\ref{fig:architecture}). The architecture prioritizes compilability, vendor\text-specific constraint enforcement, and practical deployment under industrial security requirements.

The knowledge base consolidates three retrieval segments indexed with metadata filtering: (1) function block definitions for iQ\text-R CPUs, augmented with instruction suffix semantics (\texttt{\_P} for edge execution, \texttt{\_U} for unsigned variants, \texttt{EN/ENO} conventions); (2) specification excerpts from Mitsubishi Electric manuals encoding syntax rules, reserved words, and Program Organization Unit (POU) structure constraints \cite{MEC2023,MEC2024}; and (3) auxiliary context including user\text-contributed examples and project\text-specific patterns. Documents are embedded via semantic text representations using OpenAI's 
\texttt{text-embedding-3-large} model \cite{OpenAI2024} and 
retrieved through specialized query strategies tuned for definitions, 
rule constraints, and usage examples respectively.

Prompt construction enforces vendor\text-specific hard constraints: reserved\text-word prohibition, declaration\text-before\text-use mandates, restricted datatype vocabulary aligned with GX\,Works3 semantics, and explicit POU structure templates. Retrieved context from the knowledge base is concatenated with a canonical example program encoding compliant timer usage, state\text-machine logic, and edge\text-detection idioms. 

The system supports three model configurations executed concurrently on initial queries: (1) \textbf{Azure RAG} (GPT\text-4.1 with full retrieval), (2) \textbf{Azure Standard} (GPT\text-4.1, minimal prompt, no retrieval), and (3) \textbf{Local RAG} (fine\text-tuned quantized model with retrieval). This competitive architecture enables systematic evaluation of retrieval impact and cloud versus on\text-premises inference. Follow\text-up turns use a single user\text-selected model to preserve conversational coherence. In Sec.\,\ref{sec:eval} are further models tested for evaluation purposes.

Generated code is validated via a GX\,Works3 compilation server that returns structured diagnostics. For Azure RAG responses, the system performs bounded iterative repair (maximum 3 compilation attempts). Failures trigger targeted correction prompts addressing specific diagnostic categories—undeclared variables, reserved\text-word violations, type mismatches, disallowed instructions—while preserving validated segments. The repair loop terminates on success, server timeout, or budget exhaustion.

The interface supports model selection, optional query expansion, draft mode, 
file uploads, and chat persistence. Generated responses stream with compilation results displayed in dedicated panels.

The system deploys via containerization with persistent volumes for vector storage and model artifacts. User\text-uploaded files undergo validation for binary content and encoding compliance. All proprietary code remains local; external calls are restricted to Azure OpenAI and the configured compilation server. The fine\text-tuned local model enables fully offline operation for security\text-critical environments.

In contrast to Koziolek et al. \cite{Koziolek2024b}, who demonstrated retrieval\text-augmented generation for general IEC 61131\text-3 with manual IDE import and OpenPLC validation, this architecture addresses Mitsubishi Electric\text-specific constraints through fine\text-grained retrieval segmentation (libraries/specs/general vs. unified vector store), competitive multi\text-model orchestration (three concurrent paths vs. single GPT\text-4), and automated compiler\text-driven repair loops (bounded iteration vs. manual import and spot testing). While Koziolek et al. focused on proof\text-of\text-concept with OSCAT open\text-source libraries, this work targets proprietary vendor ecosystems under low\text-data conditions via strategic prompt engineering rather than exhaustive document chunking.

Unlike approaches requiring exclusive cloud\text-dependent processing \cite{Yang2024}, extensive fine\text-tuning and generating datasets \cite{Haag2024,Ren2025}, or single\text-model architectures \cite{Liu2024}, this system operates effectively in low\text-data proprietary environments through directed retrieval, strict vendor\text-specific guardrails, and optional edge deployment.

\section{Implementation Challenges and Solutions}

The primary challenge stems from the scarcity of proprietary training data for Mitsubishi Electric's ST dialect. While official documentation exists, it comprises thousands of pages. To find corresponding descriptions, which may be partly in images, after having found the correct function block, is inefficient and error-prone. Each function block has a brief, generic description text that does not sufficiently distinguish instruction variants (e.g., \texttt{ZPUSH} vs. \texttt{ZPUSHP} for edge detection) for retrieval. We addressed this through machine-augmented descriptions: a structured prompt encoding Mitsubishi Electric-specific suffix semantics (ending in \texttt{P} for rising-edge, \texttt{\_U} for unsigned) was used to generate enhanced function block descriptions that explicitly explain variant behavior, enabling more precise retrieval.

Given industrial requirements for low latency, we opted for a comprehensive single-prompt architecture rather than multi-agent orchestration. The prompt embeds critical Mitsubishi Electric rules: reserved word blacklists, \texttt{VAR} block semantics (e.g., \texttt{VAR\_INPUT} available for \texttt{FUNCTION}), examples of incorrect label usage, and mandates that declared function blocks must be instantiated and invoked. This holistic approach performs well in a single pass.  

GX\,Works3's label-based variable management diverges standard code gerneration practices, requiring external label registration rather than inline \texttt{VAR} declarations \cite{MEC2024}. 
Generated code must reference pre-defined device labels while avoiding self-contained variable blocks that the GX\,Works3 converter rejects. The prompt explicitly instructs models to assume label pre-registration and provides templates demonstrating compliant device referencing patterns. We set up an MCP server that parses \texttt{VAR...END\_VAR} logic and imports it into GX Works 3.

Absence of existing training corpora for the local fine-tuned model necessitated synthetic data generation. We employed the same comprehensive prompt used for inference to generate task-solution pairs, then validated each via GX\,Works3 compilation. A critical validation step enforces that generated code actually invokes declared function blocks—static analysis rejects samples where function blocks are defined but unused, ensuring the compiler exercises full instruction coverage and surfaces potential errors during training data curation.

Determining optimal model configurations (cloud vs. local, RAG vs. baseline) required systematic comparison. The architecture executes three concurrent model paths—Azure RAG, Azure Standard (no retrieval), and Local RAG—on identical queries, enabling direct evaluation of retrieval impact and cloud-versus-edge trade-offs. This competitive approach surfaces performance differences in compilation success, response quality, and latency under real industrial workloads, informing deployment decisions for privacy-sensitive versus performance-critical environments.

Industrial environments often prohibit cloud dependencies due to data protection regulations \cite{Brehme2025}, necessitating on-premises inference. The local model employs quantized GGUF \cite{ggml2025} format (deepseek-coder-v2-lite fine-tuned) \cite{DeepSeekAI2024} to enable deployment on resource-constrained hardware including standard engineering laptops. Quantization reduces model size and memory footprint while preserving sufficient code generation capability for Mitsubishi Electric-specific tasks, balancing privacy requirements against computational limitations (here 10GB VRAM are sufficient).

GX\,Works3 diagnostics employ vendor\text-specific formatting that requires interpretation for automated repair. Rather than implementing complex pattern\text-based diagnostic parsing, we adopted a prompt\text-guided approach: compiler feedback is extracted verbatim from the last compilation attempt and embedded into a structured repair instruction that directs the model to address common diagnostic categories (undeclared variables, reserved\text-word violations, type mismatches, disallowed instructions) while preserving validated code segments. This approach delegates diagnostic interpretation to the LLM's reasoning capability, reducing maintenance overhead for vendor\text-specific error schemas.

These challenges underscore the viability of prompt engineering combined with strategic validation as an adaptation pathway for low-data proprietary domains. By encoding vendor-specific constraints directly into prompts, augmenting documentation with machine-generated semantic annotations, and validating synthetic data through actual compilation, the system achieves practical code generation without large-scale dataset collection or extensive model retraining \cite{Haag2024,Ren2025}. The competitive evaluation framework further enables evidence-based deployment decisions tailored to specific industrial security and performance requirements.

\section{Evaluation in Mitsubishi Electric Environment}\label{sec:eval}

\begin{table}[h]
\centering
\caption{Compilation Success Rates Across Model Configurations}
\label{tab:model-performance}
\begin{tabular}{lcc}
\toprule
Model Configuration & Compiled & Repaired \\
\midrule
Azure GPT-4.1 RAG & \textbf{73\,\%} & 23\,\% \\  
Azure GPT-4.1 Standard & 38\,\% & 5\,\% \\
Azure GPT-5 RAG & \textbf{87\,\%} & 14\,\% \\
Azure GPT-5 Standard & 56\,\% & 25\,\% \\
Local Fine-tuned RAG & \textbf{86\,\%} & 27\,\% \\
Local Fine-tuned Standard & \textbf{73\,\%} & 22\,\% \\
Local RAG & 25\,\% & 12\,\% \\
Local Standard & 23\,\% & 17\,\% \\
\bottomrule
\end{tabular}
\end{table}

With a focus on real-world applications, we incorporated extensive human feedback by releasing an early demo app for our industry partners. However, to provide a structured and empirical evaluation, we generated 100 varied queries that were used to generate 100 ST programs with each model composition. We have evaluated the four earlier mentioned models -- Azure GPT-4.1 RAG | Standard (i.e., non-RAG); Local Fine-tuned RAG | Standard (i.e., DeepSeek Coder v2 Lite Instruct) -- and further non-fine-tuned models for a broader view on our RAG approach. We got access to GPT-5 after finishing the project, but we also included it in our evaluation experiments (using reasoning set to low). All RAG configurations employ OpenAI's \texttt{text-embedding-3-large} \cite{OpenAI2024} for retrieval, ensuring consistent semantic search across model variants. The results can be found in Tab.\,\ref{tab:model-performance}.

We generated synthetic training data for the local fine-tuned model using our RAG-enhanced pipeline with GPT-4.1. To ensure diversity reflecting real-world usage patterns, we employed multiple user personas: (1)~a casual developer using informal problem descriptions, (2)~an electrician employing domain-specific jargon, (3)~a control engineer providing precise technical specifications. Task categories were systematically varied across timers, communication, array processing, pid control, etc. to cover common industrial automation scenarios.

Each generated sample underwent automated validation: first, GX\,Works3 compilation verified syntactic correctness; second, static consistency checks ensured that declared function blocks were actually invoked (rejecting samples with unused declarations) and that no forbidden identifiers or single-letter variable names appeared. Only samples passing both compilation and static validation were retained for fine-tuning. This compiler-validated filtering approach ensured training data quality without requiring manual annotation, addressing the low-data constraints inherent to proprietary industrial environments.

Our quantitative evaluation results can be found in Tab.~\ref{tab:model-performance}. As can be seen, the local fine-tuned RAG model performed best (DeepSeek Coder v2 Lite Instruct). The second best is the fine-tuned local model without RAG that achieved a 73\,\% compile rate compared to 86\% for the RAG. Interesting is the fast inference time and high quality of the local standard model, i.e. the one that did not use RAG. We also tested earlier versions of the DeepSeek model and others like CodeLlama \cite{Roziere2023} and Qwen-3 Coder \cite{Yang2025}, which performed less well. Our fine-tuned DeepSeek model was able to write code in the Mitsubishi Electric ST style and use the relevant function blocks due to training. That is, it performed well with no overhead. However, an overhead was necessary to generate curated training data. The RAG version took much longer to generate code and we had to shorten the system prompt to achieve reasonable answer times (\textless\,1\,min). 

Interestingly, the large GPT-4.1 was outperformed quantitatively by the local model. The qualitative analysis showed that users liked the larger GPT better. It delivered explanations for each generated code and also answered in text-only manner, if desired. These are things that the local model un-learned due to fine-tuning. Moreover, users told us they also liked the coding style of the larger model better: It generated longer and better structured code with various variable blocks and functions. Not all code that was compiled was reasonable, though. The smaller model often did not produce coherent logic, generated shorter programs and applied functions incorrectly. In the generated code were, e.g., empty arrays at locations that, while correct but useless, did not lead to compile errors.

However, we are comparing apples with oranges here: Models like GPT-5 or even GPT-4.1, that actually followed GPT-4.5\cite{OpenAI2025}, have surely several hundred billion parameters or more, the true number is not public, while DeepSeek Coder v2 has 16 billion \cite{GitHub.com2025}. The latter model fits into an Apple MacBooks memory, esp. when quantized, where it used about 10 GB. Hence, the users would opt for a hosted GPT-4.1/5 with RAG when data protection is not the primary concern. But, as can be seen, a local option is also available, even for highest security standards. 

From Tab.\,\ref{tab:model-performance} can be derived for sure that the fine-tuning and RAG approaches perform well, since standard models do not know Mitsubishi Electric ST style specifics. Moreover, the Azure RAGs perform pretty well in the quantitative analysis (73\%, 87\%), but convinces most in the qualitative analysis. Those generations that set up a function block or function but did not make use of it were regarded as failed attempt (not compiled). The number of repaired attempts shows that, e.g., of the 73\,\% compiled code attempts, 50\% did not need a repair, but 23\% did need up to two repairs. RAG approaches are promising; however, the latest GPT-5 performs even better while requiring fewer repairs (87\%; 14\%).

\section{Discussion and Industrial Implications}

This work demonstrates that vendor-specific coding assistants can achieve practical utility in low-data industrial domains through strategic prompt engineering and directed retrieval. Retrieval-augmented approaches substantially outperform non-RAG baselines. The 73\% success rate via fine-tuning alone indicates targeted synthetic data enables reasonable baseline performance, yet the 38\% non-RAG rate—despite GPT-4.1's size underscores the challenge of generating vendor-compliant code without domain-specific context.

Industrial practitioners prioritize data protection, security, and quality over raw performance \cite{Brehme2025}. This suggests modest performance trade-offs between cloud and local models are acceptable for eliminating external dependencies in security-critical environments. Preliminary observations show local models avoid API overhead, enabling faster iteration cycles critical for interactive development.

While this work focuses exclusively on Mitsubishi Electric's iQ-R series, the architectural components are \textit{conceptually} transferable to other ST implementations. However, we have not empirically validated cross-vendor generalization. Each platform presents distinct function block semantics and data that would require dedicated adaptation efforts. We therefore limit our claims to the Mitsubishi Electric ecosystem.

Limitations include said Mitsubishi Electric-only scope, task coverage and "specialist" local model behavior with occasional misinterpretation. Human evaluation from a single organizational context may limit generalizability; longitudinal deployment studies are needed to assess adoption barriers and real-world productivity impacts. The local model limits the available context and thus shortens the system prompt and synthetic training data may have downsides not yet visible.

We acknowledge that compilation success is a necessary but not sufficient criterion for code quality. Functional correctness would require domain-expert evaluation or execution on physical hardware, which we identify as a limitation of this study even though we let expert users validate compilation results qualitatively. Future work should incorporate simulation-based testing or hardware-in-the-loop validation to assess runtime behavior.

Automatically generated PLC code for industrial control systems poses inherent risks that must be addressed in deployment contexts. Faulty control logic could lead to equipment damage, production downtime, or safety hazards for personnel. We emphasize that our system is designed as an \textit{assistive} tool, not a replacement for human verification. All generated code must undergo a mandatory review by qualified automation engineers before 
deployment or simulation testing in vendor-provided environments. Moreover, it must undergo compliance verification against site-specific safety standards (e.g., cyber security).

Deployment-wise, containerized on-premises configurations with local models enable SME adoption without cloud dependencies. Integration into GX\,Works3 (analogous to other vendors) would reduce context-switching overhead. Synthetic training data—validated via compiler—enables continuous improvement, while customer-specific uploads tailor the system to organizational conventions without external data exposure, addressing industrial security requirements.

Prompt engineering is cost-effective for low-data domains, avoiding multi-month fine-tuning campaigns. Future work includes static pre-compilation validation, coding environment integration, but mostly a multi-agent approach for increased performance in terms of quality and velocity. We also plan to employ our findings to graphical IEC 61131-3 languages. Our aim is to support the whole development cycle.

\bibliographystyle{IEEEtran}
\bibliography{generist.bib}
\end{document}